\documentclass[10pt]{article}
\usepackage[dvips]{epsfig}
\usepackage[T1]{fontenc}
\usepackage[latin1]{inputenc}
\usepackage{graphicx}
\usepackage[english]{babel}
\usepackage{amsmath}
\usepackage{amssymb}
\usepackage{amsfonts}
\usepackage[T1]{fontenc}
\usepackage{color}
\usepackage{babel}
\usepackage{verbatim}
\usepackage[unicode=true,pdfusetitle,bookmarks=true,bookmarksnumbered=false,bookmarksopen=false,
 breaklinks=false,pdfborder={0 0 1},backref=false,colorlinks=true]{hyperref}
\hypersetup{linkcolor=blue,citecolor=blue}
\makeatletter
\usepackage[dvips]{epsfig}
\usepackage[T1]{fontenc}
\usepackage{color}
\usepackage{pdflscape}

\begin{document}

\title{\bf D-bound and Bekenstein bound for McVittie solution  surrounded
by dark energy cosmological fields    }
\author{H. Hadi$^{a}$\thanks{%
email: hamedhadi1388@gmail.com}, Y. Heydarzade$^{b}$\thanks{%
email: yheydarzade@bilkent.edu.tr},  F. Darabi$^{\star~a}$\thanks{%
email: f.darabi@azaruniv.ac.ir, Corresponding author} ~~and K. Atazadeh$^{a}$\thanks{%
email: atazadeh@azaruniv.ac.ir}   
\\{\small $^{a}$Department of Physics, Azarbaijan Shahid Madani University, Tabriz,
Iran}
\\{\small $^{b}$Department of Mathematics, Faculty of Sciences, Bilkent University, 06800 Ankara, Turkey}}
\date{\today} 

\maketitle
\begin{abstract}
 The cosmological candidate
fields for dark energy  as
quintessence, phantom and cosmological constant, are studied in terms of an entropic hypothesis  imposed on the McVittie solution surrounded by dark energy. We certify this hypothesis
as    ``$D$-bound-Bekenstein bound identification'' for  dilute
systems and use it as a criterion to determine which candidate of dark energy
can satisfy this criterion for a dilute McVittie solution. It turns out that  only the cosmological constant can pass this criterion successfully  while  the quintessence and phantom fields fail, as non-viable dark energy fields for this particular
black hole solution. Moreover,
assuming this  black hole to possess the saturated entropy,  the entropy-area law  and the holographic principle can put two constraints on the radius $R$ of the cosmological horizon.  The first one shows that the Hubble radius is discrete such that for any arbitrary value of the black hole mass $m_{0}$, the value of $R$ is determined up to an integer number. The latter one shows that when a black hole is immersed in a cosmological background, the radius of  the cosmological horizon is constrained as $R<\frac{1}{H}$.  
\end{abstract}

 \section{Introduction}
 The cosmological constant  is the simplest candidate for dark energy to describe the current accelerating expansion of
the universe \cite{AU}. The corresponding $\Lambda CDM$ model
is in very good agreement  with observations. However, $\Lambda CDM$ model suffers from two well known ``cosmological constant problem'' \cite{CCP} and ``coincidence problem'' \cite{CP}. Other alternative models of dark energy have then introduced  such as  quintessence  model  \cite{Caldwell}, with a canonical scalar field having a particular potential
in the presence
of matter and radiation which could interpret the late-time cosmic acceleration 
 \cite{fujii,ford,wett,ratra}.  The other alternative
for dark energy is phantom model with a phantom scalar field rolling up the potential because
of the negative kinetic energy \cite{shin1,shin2,shin3,carroll12,singh12,sami12}.  The equation of state parameter of the phantom field in the case of an exponential
potential  results
in a big-rip singularity which is considered as a problem. To study more
about  weirdness
of phantom and quintessence fields refer to \cite{phys1,phys2,phys3,phys4,phys5,phys6,phys7,hadi1,
hadi2}.  In
addition, the strange behaviour  of  quintessence and phantom fields in a
particular interval of time, in this paper, 
is in agreement with the problematic constraint on the quintessence energy density in the
early cosmological era and the late time cosmological behaviour of phantom field  leading to a big rip singularity \cite{ratra,cline}.  Quintessence potentials are classified into \textquotedblleft freezing models\textquotedblright\
 and \textquotedblleft thawing models\textquotedblright\ \cite{linder}. These potentials are
imposed by particular features at early universe in order to have quintessence
as dark energy source in late
time universe \cite{ratra,zlatev,models,brax,linde1,linde2,linde3}. On the
other hand, in order
to have a consistent phantom
scalar field theory which leads to big rip singularity in the late time universe, one needs to go
beyond the framework of particle physics \cite{carroll,cline,singh,sami,gann}. 
 
In this paper, using a thermodynamical approach,
we intend to study the McVittie solution surrounded by
cosmological constant, phantom
and quintessence fields
 and show that the entropy bounds can exclude phantom
and quintessence fields from being considered as dark energy candidates  for
this solution.
The  use of  some relevant entropic limits
can be considered as a  powerful thermodynamical approach in the study of dark energy models. In fact, the  equations of motion can perfectly   predict
the time-reversible behavior of  dynamical  systems, nevertheless, for thermodynamical systems
the time-reversibility  is not guaranteed  due to some entropic
consideration.  Dynamical black holes, surrounded by the cosmological fields, are considered as such   thermodynamical systems and one can use the relevant entropic considerations to study about the viability of these surrounding
cosmological fields from thermodynamical point of view.  In this regard, we  study the entropic bounds in the McVittie  solution surrounded by
cosmological  constant, phantom and quintessence fields   and try to find a thermodynamical
criterion by which one can single out  the  viable dark energy field that can satisfy
this thermodynamical
criterion.

The McVittie solution \cite{mass}
is one of the exact  solutions of Einstein's general theory
of relativity which describes a  black hole evolving in time. This dynamical
solution
appears on a wide
range of problems stemming from the perfect-fluid cosmology
to the scalar field actions and modified theories of gravity \cite{valeri}. This solution has the following properties:  $i)$ it is a spherically symmetric,
shear free and perfect fluid solution of the
Einstein field equations, $ii)$ the near field is the Schwarzschild solution
in the isotropic coordinates with the mass parameter $m$, and  $iii)$ the far field limit is the Friedmann-Robertson-Walker (FRW)
spacetime with the scale factor $a(t)$. 
 Then, by setting $m=0$ in the metric, one find FRW metric as the background,
and by setting $a(t)=1$ one
 arrives at the Schwarzschild solution, 
in the isotropic coordinates, as the local inhomogeneity. However, the global structure of McVittie spacetime has been a prominent issue for the
controversial question of whether this solution describes a black hole 
embedded
in an isotropic FRW universe or not \cite{nol, far}.   The answer to this question has been
found recently in  \cite{kaloper, lake}. In \cite{silva, noli}, it is discussed
that  how the particle motion and the
structure of the event horizon in this solution ultimately depend on
the cosmological history.
For an expanding FRW background, the McVittie solution possesses  a weak
singularity  at $\bar r = m/2$  which is a spacelike hypersurface  and the
past boundary of spacetime \cite{nol}. This locus manifests itself as a part of future boundary of spacetime when 
there exists a non-negative cosmological constant. For the de Sitter background the singularity  $\bar r = m/2$  disappears and its 2-surface describes a true black hole horizon. To clarify more on these
features, the McVittie solution has been generalized to the spatially non-flat
backgrounds \cite{global}.
As the generalizations of the McVittie solution, there are also solutions representing
the charged black holes in an expanding FRW universe \cite{faro, zan, vaid}, see also \cite{davidson2} for the general relativistic perfect fluid black hole solutions and \cite{met, kra} for the classification and other possible generalizations.
In derivation of the general
solution in \cite{met} (including the McVittie
solution), the scale factor $a(t)$  arises as an integration constant when one integrate the field equations,   see the sections II and III of  \cite{met} where $q(t)=a(t)/\dot a(t)$ and the energy-momentum is  a perfect
fluid. Then, to determine the arbitrary scale factor $a(t)$,
 one possibility is to consider an equation of state for the matter field
 supporting the Einstein field equations as it is usually done for the Friedmann
equations, i.e $p=\omega \rho$, or make an ansatz for its form  for various cosmological eras according to the standard model of cosmology. In the case of McVittie solution describing an inhomogeneity in FLRW background \cite{kaloper,lake}, the perfect fluid source represents the matter field filling the background FLRW universe obeying the Friedmann equations in the cosmic scale.
Thus, this cosmic fluid can be radiation, dust, quintessence, cosmological constant or a phantom field regarding its equation of state. However,
there has yet been no proof that a McVittie metric with a dark energy equation of state other than that of a cosmological constant (i.e. for $\omega\neq -1$)  represents a true black hole that contains an event horizon. In the
literature, what is mainly studied is the properties of apparent horizons.
As an instance,   the apparent horizons for the McVittie solution in a phantom background is discussed in \cite{far}. For more discussion on McVittie solution in phantom and quintessence backgrounds, one may also see \cite{5,6}.

In this paper, motivated by the fact that the McVittie solution may represent
 a dynamical black hole in a cosmic background and then it must obey the
thermodynamical
 entropy constraints, we put this solution under the  scrutiny of the D-bound \cite{boss1} and Bekenstein bound \cite{bek1, bek2, bek3}. We
obtain these entropy bounds for the McVittie solution in a general form.  Then, we discuss about the dilute system limit of these bounds for various
cosmological background fields, those are the solutions to the Friedmann
equations, and noting that
both these bounds are  direct results of the generalized second law of
thermodynamics (GSL), we demand their identification in the dilute system limit, as a hypothesis of ``$D$-bound-Bekenstein bound identification'' \cite{boss1}.
Here, we note  that because
the entropy bounds deal with the cosmological apparent horizon and  the considered cosmological fields solves the standard Friedmann equations at this cosmological scale, then the McVittie solution and its entropic considerations are meaningful.  
We show  that these entropy bounds put some restrictions on the parameters of the McVittie solution
with different cosmological background fields. In the following we review
these entropy bounds which are essential for our purpose. 

The Bekenstein bound claims that there is 
an upper bound  on the entropy of any physical system.
It states that for an isolated and stable thermodynamic system within
an asymptotically
flat space, there is a constraint on the universal entropy of the system as
follows 
\begin{equation}\label{master}
S_{m}\leqslant 2\pi RE,
\end{equation}
where $E$ is the total energy and $R$ is the radius of the system. Bekenstein bound has been considered in two forms, the empirical from \cite{bek3, NNm,
pag,pag*,wald*}
and the logical form  \cite{bek1, bek2, bek6, bek7}. Its logical form is based on the generalized second law of thermodynamics (GSL) and the Geroch process.  To study about   quantum effects on this bound, one can
refer to \cite{un1,un2,pel}.

D-bound has been introduced by Bousso  \cite{boss1} and investigates a gravitational system by a gedanken experiment as follows. Suppose a black hole in the universe with a cosmological horizon. The total entropy of the system in a case when there is a matter system (a black hole) and a cosmological horizon reads as $S_{m}+S_{h}$ where
$S_m$ and $S_h=A_c /4$ are the entropy of the matter system and the cosmological horizon with the area $A_c$,
respectively. Now, suppose that  an
observer moves away from the matter system until the matter system falls out
of the apparent
horizon of this observer. This observer is a witness of crossing the matter
system from the apparent
horizon in a thermodynamical process. Then, the total entropy of the final system
 is given
by $S_{0}=\frac{A_{0}}{4}$ where $A_0$ is the horizon area in the absence
of the matter system. Now,  According to the GSL, the observer can put an
upper bound to the entropy of the matter system by
 \begin{equation}\label{entropy bound}
S_{m}\leqslant \frac{1}{4}(A_{0}-A_{c}).
\end{equation}
This  is the D-bound on the matter system which has been introduced  firstly for the de Sitter space. The identification of the D-bound and  the Bekenstein bound on the matter system has been shown
for the de Sitter background.  Also, it has been indicated that this identification can be generalized to the arbitrary dimensions \cite{boss1}.
Following the same method in \cite{boss1},  the  D-bound entropy for the various possible black hole solutions on a four dimensional brane, and its identification with the Bekenstein bound
have been investigated in \cite{DDDbek}. It was shown that there are some differences in the D-bound
entropy for the solutions on a brane within a higher dimensional bulk in comparison to the usual four dimensional black hole solutions.
It is concluded  that   these differences are because of the extra loss of information
due to the extra dimensions when black hole is crossing from the  apparent horizon of the observer confined to the four dimensional brane. Bearing these
in mind, it seems that the hypothesis of identification of D-bound with  Bekenstein bound is a reasonable ingredient in a thermodynamic treatment
of Black holes, with entropic considerations. Finally,
we assert that since the
D-bound is a direct result of GSL, it is also definable   for
 a dynamical system like  McVittie solution having rapidly evolving matter fields and
causal horizon  \cite{wallll, tjackkk}.     

The organization of this paper is as follows. In section 2, we obtain the D-bound and Bekenstein bound for McVittie Solution in a  general form.
Then, we discuss about the dilute system limit for different cosmological backgrounds as the candidates for  dark energy. In section 3,
using the entropy-area law and the holographic principle, we obtain two constraints
on the cosmological horizon radius. 
The paper ends with a conclusion in section 5.
\section{D-bound and Bekenstein bound for McVittie Solution }
In this section, we study the McVittie solution \cite{mass}  and derive the
corresponding D-bound and Bekenstein bound.
The McVittie solution represents the embedding of the Schwarzschild solution
within the  FRW cosmological background which  can be generally de Sitter or not. The line element of this solution in the isotropic
coordinates reads as
\begin{equation}\label{mcv}
ds^{2}=-\frac{(1-\frac{m(t)}{2 \bar r})^{2}}{(1+\frac{m(t)}{2 \bar r})^{2}}dt^{2}+a(t)^{2}(1+\frac{m(t)}{2
\bar r})^{4}(d \bar r^{2}+ \bar r^{2}d\Omega^{2}_{2}).
\end{equation}
Considering  $G_{\mu \nu}$ as  the Einstein tensor, the McVittie no-accretion condition is given by ${G^{\bar r}}_{t}=0={T^{\bar r}}_{t}$ which forbids the accretion of the cosmic fluid onto the
central black hole. This condition   leads to the  following differential equation 
\begin{equation}\label{flow}
\frac{\dot m}{m}+\frac{\dot a}{a}=0,
\end{equation}
which can be  solved as
\begin{equation}
m(t)=\frac{m_{0}}{a(t)},
\end{equation}
where $m_{0}$ is a positive constant and $a(0)=1$\footnote{In
the modern language,  the equation (\ref{flow}) represents  the constancy of the Hawking-Hayward mass $m_H$, i.e. $\dot m_H=0$ \cite{hayw}. Indeed, $m(t)$ here stands just
as a metric coefficient in a particular coordinate system. In fact,  one has
to identify the Hawking-Hayward mass $m_H$ \cite{hayw} as the physically relevant mass, which eventually is related to   the physical size of the central
object or  its corresponding horizon, in order to avoid making any coordinate-dependent
statements on the mass and size \cite{zan, hayw}
or the temperature \cite{temp} of the central object. }.
The metric (\ref{mcv}) can be written in the Schwarzschild coordinates for
more convenience. To
do this, one can define the areal radius $R$ as \cite{vale}
\begin{equation}\label{areal}
R=a(t)\bar r\left(1+\frac{m(t)}{2\bar
r}\right)^{2},
\end{equation}
and finds the identity 
\begin{equation}\label{iden}
\frac{(1-\frac{m(t)}{2\bar r})^{2}}{(1+\frac{m(t)}{2\bar r})^{2}}= 1-\frac{2m_{0}}{R}.
\end{equation}
The  equation (\ref{flow}) gives
\begin{equation}\label{H*}
H\left( 1+\frac{m}{2\bar r} \right)+ \frac{\dot m}{\bar r}=H\left(1-\frac{m}{2\bar
r} \right)
\end{equation}
where $H=\dot a/ a$ is the Hubble parameter of the FRW background.
Then, using (\ref{areal}), (\ref{iden}) and (\ref{H*}) the metric can be written as 
\begin{equation}\label{metriccc}
ds^{2}=-(1-\frac{2m_{0}}{R}-H^{2}R^{2})dt^{2}+\frac{dR^{2}}{1-\frac{2m_{0}}{R}}-\frac{2HR}{\sqrt{1-\frac{2m_{0}}{R}}}dt
dR+R^{2}d\Omega^{2}_{2}.
\end{equation}
To remove the cross term,  we use  the coordinate transformation
\begin{equation}
dT=\frac{1}{F}(dt+\beta dR),
\end{equation}
where $F(t,R)$ is an integration factor 
 for the above closed 1-from and $\beta(t,R)$  ensures that in the new coordinate
system, the $dtdR$ component of the metric vanishes.  The function $\beta(t,R)$  can be found as \cite{vale}
\begin{equation}
\beta(t,R)=\frac{HR}{\sqrt{1-\frac{2m_0}{R}}\left(1-\frac{2m_0}{R}-H^2 R^2\right)}.
\end{equation}
Then, our line element (\ref{mcv}) becomes
\begin{equation}\label{metricc}
ds^{2}=-(1-\frac{2m_{0}}{R}-H^{2}R^{2})F^{2}dT^{2}+\frac{dR^{2}}{1-\frac{2m_{0}}{R}-H^{2}R^{2}}+R^{2}d\Omega^{2}_{2}.
\end{equation}
For  $H= constant$, the integration factor $F(t,R)$ can be set to unity, and then this metric reduces to the Schwarzschild-de Sitter metric. 

For this spacetime, the Misner-Sharp-Hernandez mass
$M_{MSH}$ \cite{mis} representing the total mass within the radius $R$  can be found as

\begin{equation}\label{msh}
M_{MSH}=\frac{4\pi G}{3}\rho R^{3}+m_{0},
\end{equation}
where $\rho(t)=\frac{3}{8\pi}H^{2}(t)$ is the background cosmic fluid density. Thus,  $M_{MSH}$ includes both the energy of the background cosmic fluid inside the sphere of radius $R$  and the mass of the local inhomogeneity $m_{0}$. 

To obtain the D-bound on the matter system in McVittie spacetime, we need to derive the physical
horizons of the system, one as the inner horizon,
which is an anti-trapping surface representing an inner apparent horizon
\cite{78}, and the other one as the cosmological apparent horizon. The  locations
of the horizons for the metric (\ref{metricc}) are given by $g^{RR}=0$ which
leads to

\begin{equation}
H(t)^{2}R^{3}-R+2m_{0}=0.
\end{equation}
This equation has the following solutions
\begin{equation}\label{r1}
R_{1}=\frac{2}{\sqrt{3}H}\sin(\psi),
\end{equation}
\begin{equation}\label{r2}
R_{2}=\frac{1}{H}\cos(\psi)-\frac{1}{\sqrt{3}H}\sin(\psi),
\end{equation}
\begin{equation}
R_{3}=-\frac{1}{H}\cos(\psi)-\frac{1}{\sqrt{3}H}\sin(\psi),
\end{equation}
where $\sin(3\psi)= 3\sqrt{3}m_{0}H$. The solution $R_{3}$ is nonphysical because for an expanding universe with the positive $H$, $R_{3}$ becomes negative. There are strong evidences that, in the absence of event horizons, one can ascribe an entropy to the apparent horizons. The thermodynamic behaviour of these horizons has been considered extensively \cite{apparent}. Therefore, using the Bekenstein-Hawking entropy-area law $S=A/4$, we ascribe
the entropy to the apparent horizons  $R_{1}$ and $R_{2}$ as $\pi
R_{1}^{2}$ and $\pi
R_{2}^{2}$, respectively.

   To derive D-bound  ($\ref{entropy bound}$) for the McVittie solution, we consider  the radius $r_0$ of the system as $r_{0}=R_{2}|_{m_{0}=0}=1/H$, when the mass-energy of the system  is only due to the   cosmic fluid ($\rho(t)$) inside
the system. In the presence of the matter inhomogeneity $m_0$ as well as the
cosmic fluid, the radius of the cosmological horizon is $r_{c}=R_{2}$
.  Therefore, the initial entropy of the system is $S_{m}+S_{\rho}+\frac{A_{c}}{4}$
and  the final entropy of the system becomes $S_{\rho}+\frac{A_{0}}{4}$. Thus, using the Bekenstein-Hawking entropy-area law, the D-bound (\ref{entropy bound}) for the matter system of the McVittie solution reads as
\begin{equation}\label{bekkkk}
S_{m}\leqslant \pi(r_{0}^{2}-r_{c}^{2}),
\end{equation}
where, substituting $r_0$ and $r_c$ using (\ref{r2}), it takes the following
 form
\begin{equation}\label{dmc}
S_{m}\leqslant \pi(\frac{2}{3H^{2}}\sin^{2}\psi +\frac{2}{\sqrt{3}H^{2}}\cos
\psi
\sin \psi).
\end{equation}
This bound is the D-bound for the McVittie solution for the generic backgrounds.

For the dilute system limit \cite{boss1}, the size of the local inhomogeneity
$R_1$
  must be negligible
in comparison to the cosmic  horizon $R_2$. Indeed, this means that the local
mass $m_0$ should be very small relative to $H^{-1}$ indicating the cosmic Hubble horizon
radius. Then,  regarding (\ref{r1}) and (\ref{r2}),  one realizes that the dilute system
limit is equivalent to the approximation relations  $\sin\psi\approx \psi$
and $\cos\psi \approx 1$ where  $\psi = \sqrt{3}m_{0}H$. Using these relations,
we obtain $R_1\approx 2m_0$ and $R_2\approx H^{-1} - m_0\approx H^{-1}$ which indicate
the approximate Schwarzschild and Hubble horizon radii, respectively. Therefore, the general D-bound (\ref{dmc})  reduces to
\begin{equation}\label{17}
S_{m}\leqslant \frac{2\pi m_{0}}{H}, 
\end{equation}
for the dilute systems.

Now, our aim is to obtain the Bekenstein bound for the McVittie solution. To
do this, we  follow Bousso \cite{boss1} for the definition of the Bekenstein bound  as 
 
\begin{equation}\label{bekenflat}
S_{m}\leqslant \pi r_{g}R,
\end {equation}
where $r_{g}$ is the gravitational radius of the matter system and $R$ is
the circumscribing radius of the system. For the McVittie solution these
radii correspond
to $R_1$ and $R_2$ in (\ref{r1}) and (\ref{r2}), respectively. Then, the
 Bekenstein bound reads as
 \begin{equation}\label{bmc}
S_{m}\leqslant\frac{2\pi}{\sqrt{3}H^{2}}\cos\psi
\sin\psi-\frac{2\pi}{3H^{2}}\sin^{2}\psi.
\end{equation}
This form is the general form of the Bekenstein bound for the
McVittie spacetime. Then, one realizes that in the general cosmological setup
of  McVittie solution, the D-bound and the Bekenstein bound do not coincide
and indeed the latter is  tighter. However, using the dilute system  conditions, as discussed  after
Eq.(\ref{dmc}), the above Bekenstein  bound reduces to  the form  $S_{m}\leqslant \frac{2\pi m_{0}}{H}$. 

 We summarize our findings till now in the following
remark. \\
\\\textbf{Remark 1:} \textit{The entropy D-bound and the Bekenstein bound do not coincide for  a  generic cosmological background in McVittie spacetime. 
Indeed, the latter bound is  always tighter. However, for any general cosmological background
satisfying the ``dilute system limits'', these two bounds are identified
in McVittie spacetime.}\\

In the following subsections, we investigate these entropy bounds for
the de Sitter, phantom and quintessence backgrounds in more details. 
\subsection{McVittie Solution in de Sitter Background and Entropy Bounds}
 
 Here, we consider the de Sitter background in which $H=\mbox{constant}\propto \sqrt \Lambda$ where $\Lambda$ is the cosmological constant. Since the value of $\Lambda$ is constant and very small, then
one can always use the dilute system limit for the de Sitter background \cite{boss1}
which
results in the coincidence of the D-bound and the Bekenstein bound. The interesting point about the cosmological constant is
that the dilute system limit (i.e. $m_0<<1/H$) is provided for both the early and
late times.   This turns out that the cosmological constant is an special fluid providing the  coincidence of the D-bound and the Bekenstein bound
for all cosmic times. In the following subsections, we show that this identification
does not happen
in general  for the phantom and quintessence fields for all cosmological time intervals.  
  
\subsection{McVittie Solution in Phantom Background and Entropy Bounds}
The McVittie solution in the phantom background with the barotropic  equation of state $\omega <-1$ have been studied in \cite{valij, cal*}. A phantom dominated universe evolves toward a finite time big rip singularity
\cite{sing} as
\begin{equation}
a(t)=\frac{a_{0}}{(t_{rip}-t)^{\frac{2}{3|\omega+1| }}}
\end{equation}
where $a_{0}$ is a constant. The derivation of Hubble parameter is straightforward
and concisely is given by 

\begin{equation}\label{H**}
H(t)=\frac{2}{3|\omega+1|}\frac{1}{t_{rip}-t}.
\end{equation}
 For the phantom background, the D-bound is the same as the general equation ($\ref{dmc}$) where the Hubble parameter
is replaced by (\ref{H**}). Regarding (\ref{r1}), (\ref{r2}) and (\ref{H**}), one can discuss about the dilute system limit and consequently the coincidence of
the D-bound and the Bekenstein bound
for a phantom background  as follows.
\begin{itemize}
\item At the early
times, both the local apparent  horizon $R_1$ and the cosmological horizon $R_2$ exist, and they  are located approximately at $2m_0$ and $1/H$, respectively.
At the early times when the phantom field is not the dominant field, the dilute system limit is provided. This is supported by the fact
that for small $t$ values, and since big rip time $t_{rip}$   is large enough, then
$H$ is small and consequently the approximation relation $\sin(\psi)\approx\psi$
is valid. 
 \item As time progresses, the phantom field dominates the cosmos such that $H$ and consequently $\rho (t)$
grow up and diverge at the finite time big rip state. Then, 
the dilute system approximation fails for the late time cosmology when the phantom field dominates the universe.
\end{itemize}
 The consequence of the above two points is that for the McVittie solution in the phantom background, the coincidence of the obtained general D-bound (\ref{dmc}) and the Bekenstein bound (\ref{bmc}) happens only for the early times when the phantom field is not the dominant field in the universe and consequently the dilute system limit is provided.
As the cosmic time progresses and the phantom field dominates the universe,
the dilute system limit fails and these two bounds
deviate. This behavior is one of the  weird features of the cosmological
phantom fluid, which also leads to the violation of the second law of
thermodynamics in many ways \cite{vio}.
Here, one notes that the asymptotic behavior of McVittie horizons is far from trivial and differs significantly from an FLRW solution in extreme regimes, see \cite{78} where the effect of the cosmological expansion, for the different
choices of scale factor, on the causal structure of the McVittie solution is studied.

\subsection{McVittie Solution in Quintessence Background and Entropy Bounds}
For the quintessence background with $-1<\omega<-1/3$, the scale factor of
the FRW universe is given
by \cite{sing} 
\begin{equation}
a(t)=a_{0}t^{\frac{2}{3(\omega +1)}},
\end{equation}
where $a_{0}$ is a constant. Thus, the Hubble parameter reads as 

\begin{equation}\label{H}
H(t)=\frac{2}{3(\omega +1)}\frac{1}{t}.
\end{equation}
Here, regarding (\ref{r1}), (\ref{r2}) and (\ref{H}),  one notes to the following points.
\begin{itemize}
\item At the early times, the dilute system limit, $m_0<<1/H$,  fails for the solution
(\ref{H}) which represents  divergent $H$ and $\rho(t)$ for a quintessence
field.  Generally, quintessence potentials are classified into \textquotedblleft freezing models\textquotedblright\
 and \textquotedblleft thawing models\textquotedblright\ \cite{linder}. These potentials are
imposed by particular features at early universe in order to have quintessence
as dark energy source in late
time universe \cite{ratra,zlatev,models,brax,linde1,linde2,linde3}. 
 \item For the late times, $H$ decreases and the dilute system limit $m_0<<1/H$
 is provided.
\end{itemize}
In contrast to the phantom field, for a quintessence field the dilute system limit
is provided only for the late times, and consequently the coincidence of the D-bound (\ref{dmc})
and the Bekenstein bound (\ref{bmc})
happens only for the late times when the quintessence appears as the dark
energy source of the expansion of the universe.\\
  
We conclude our findings in these three subsections in the following remark. \\
\\
\textbf{Remark 2:} \textit{Derivation of the entropy D-bound and Bekenstein
bound for the McVittie spacetime turns out that the cosmological
constant field with $\omega =-1$ is the unique cosmological field which provides the identification of the D-bound and  the Bekenstein bound for all cosmic
time intervals. Any deviation
from the barotropic equation of state parameter $\omega =-1$ perturbs this identification. For the phantom field possessing $\omega <-1$, the identification of these two bounds exists only at the early times and is lost for the late times. For
the quintessence field with $-1<\omega<-1/3$, there is no identification for the early universe but as the universe expands this identification emerges.
Then, entropy criteria for McVittie spacetime imply the cosmological constant as the most viable dark energy candidate.}\\

Before ending this section, it is worth mentioning that there are other matter fields like radiation and dark matter playing a role in the McVittie solution
 at the early times \cite{kaloper}. However, regarding  our aim in this paper which is finding an entropic criterion on dark energy fields by comparing
the the D-bound  with Bekenstein bound, the existence of a cosmological horizon is essential  for constructing D-bound. Indeed,   the existence, location and behavior of  horizons at early times are the essential requirements to construct the D-bound and discuss  its identification with Bekenstein bound. One notes that the existences of the inner and outer apparent horizons of McVittie solution given in (\ref{r1}) and (\ref{r2}), respectively,
  are subjected to the condition $0<\sin(\psi)<1$ or $m_{0}H(t)< \frac{1}{3\sqrt{3}}$
  \cite{far}. Therefore, there is a critical time
for the existence of an apparent horizon
for any background field determined  by $m_{0}H(t_{c})=\frac{1}{3\sqrt{3}}$. As an instance, for a dust background with  the Hubble parameter $H(t)=\frac{2}{3t}$, this critical time is $t_{c}=2\sqrt{3}m_{0}$.
 Then, in the context of McVittie solution, there is no  physical apparent horizon at the early times, i.e for  $t<t_{c}=2\sqrt{3}m_{0}$, 
 in a dust-dominated background. This fact is also addressed in detail in
 \cite{far},  and one can observe from the figure 1  in \cite{far} that at the early times none of the inner and cosmological apparent horizons appear.
  Then,  the construction of  D-bound fails before the critical time $t_{c}=2\sqrt{3}m_{0}$  when there is no cosmological horizon.  The same argument  applies for the radiation field possessing the Hubble
parameter  $H(t)= \frac{1}{2t}$.  However,
the situation changes  when there is a phantom filed in the background.  For a phantom background, the Hubble
parameter reads as $H(t)=\frac{2}{3|\omega+1|}\frac{1}{t_{rip}-t}$ and the time interval  which  the apparent horizons exist reads as  $t<t_{rip}-\frac{2\sqrt{3}m_{0}}{|\omega+1|}$
where $t_{rip}$ is the future cosmological big rip time.
Hence, when the cosmological background possesses a phantom field, there are both the cosmological and local horizons in early times and one can
construct D-bound and investigate its identification with Bekenstein bound. This point can  be seen in the figure 3 of 
\cite{far}  where both the inner and cosmological apparent horizons exist before $t_{rip}$.
   According to these considerations, we have addressed
the phantom background at the early times. However,  one notes that   in the early times, the total dynamics of the McVittie solution are complicated, and it is governed by the total matter source that includes dark matter, radiation, and dark energy sources in general.  

\section{Constraint on the radius of cosmological horizon in McVittie Spacetime }
In this section, we use discrete behaviour of entropy-area law  to derive constraint on the radius
 of cosmological horizon in McVittie solution
surrounded by cosmological fields. In the first subsection, we consider the discrete behaviour of entropy-area law and in the next subsection
we attempt to put a constraint on the radius of  cosmological horizon by Holographic
principle and maximum entropy of the whole system.  \subsection{Discrete behaviour of  Entropy-Area law and radius of cosmological horizon}
\subsubsection{The Case of de Sitter Background ($\omega =-1$) }
 As we mentioned in the section
2.1, the D-bound or Bekenstein bound for $\omega =-1$ read as $S_{m}\leqslant \frac{2\pi m_{0}}{H}$ for generic matter systems. For a black hole which
possesses the saturated entropy, we have 
\begin{equation}\label{quan}
S_{m}=\frac{2\pi m_{0}}{H}.
\end{equation}
On the other hand, by putting the value of the entropy of  black
hole $S_{m}=\frac{A}{4l_{p}^2}$ in equation ($\ref{quan}$), we obtain\footnote{Note that in all of the other formulas we have put $l_{p}=1$ but for the clarification
here we have inserted the $l_{p}$ in equation (\ref{max}).}
\begin{equation}\label{max}
\frac{2\pi m_{0}}{H}=\frac{A}{4l_{p}^2}=\frac{\alpha N}{4},
\end{equation}
where we have put $A= \alpha l_{p}^2N$ as the area of  black hole and $N$ is an integer number.  $\alpha$ is the proportionality constant
and is an
${\cal O} (1)$ dimensionless coefficient \footnote{There is also an interesting method for calculation of the entropy 
of black hole  by the method of graph theory  \cite{Davidson1}.}. The proportionality constant $\alpha$
has been considered by various values. The requirement which demands the 
number of state $e^{S}$ to be integer, leads to $\alpha=4 \ln q$ where q is an integer \cite{spec}. 
   Various kinds of arguments impose  different appropriate integers, such
as $q = 2, 3$ \cite{are,prin,quasi}. Consistent highly damped quasinormal modes (QNMs)  \cite{mode}   
  and "holographic shell model" for BHs demand $\alpha=8\pi$ \cite{inter,area,kerr} and $\alpha = 8 \ln 2$ \cite{stack}, respectively. Finally, by considering
all of these constraints, the interval  $1 < \alpha < 30$ is a consistent
and reasonable range  \cite{spec,are,prin,quasi,mode, inter,area,kerr}.   Now, with respect to equation (\ref{max}), the Hubble radius $R=1/H$ is given by 
   \begin{equation}\label{R}
R=\frac{\alpha }{8\pi m_{0}}N,
\end{equation}
which shows that the Hubble radius is discrete such that for any arbitrary value of $m_{0}$, the value of $R$
is given up to an integer number.
\subsubsection{The Case of  Background with $\omega \neq -1$}
 In the phantom background, assuming that the matter system is a black hole, the Bekenstein entropy  becomes
 \begin{equation}\label{sinn}
S_{m}= \frac{2\pi}{\sqrt{3}H^{2}}\cos \psi
\sin\psi-\frac{2\pi}{3H^{2}}\sin^{2}\psi.
\end{equation}
By putting the value of the black hole's entropy
$S_{m}=\frac{A}{4l_{p}^2}$ in equation ($\ref{sinn}$)
we obtain
\begin{equation}
\frac{2\pi}{\sqrt{3}H^{2}}\cos\psi
\sin\psi-\frac{2\pi}{3H^{2}}\sin^{2}\psi=\frac{A}{4l_{p}^2}=\frac{\alpha N}{4}.
\end{equation}
 Here, unlike the previous case, we are not able to write  $H$ up to a  discrete
number for the late times. This is  because of that $\sin(\psi) $ and $\cos (\psi) $ terms have nonlinear dependence on $H$ and
$m_0$.
Therefore,   in this case, the radius of the cosmological horizon in the presence of black hole does not show any discrete behaviour in late times, but for early
times the value of $H$ in (\ref{H**}) is small and as we mentioned in subsection
2.2 the system can be diluted we can consider this case as a dilute
case (\ref{quan}), so the interpretations are like as $\omega=-1$ case.

In the quintessence background, we have the same interpretation  as the cosmological constant phase $\omega=-1$ for the late times, but for the early times the system
cannot be dilute. Then, the radius of cosmological horizon in the presence of black hole does not show any discrete behaviour for the early times. 
\subsection{Constraints via the Holographic Principle}
 One can think of maximum entropy of the system in McVittie spacetime. The maximum entropy of the system of black hole and FLRW background with a cosmological horizon is $\pi R^{2}$ and $R$ is the system radius which is the radius of
cosmological horizon $R_{2}$. Now, according to the Bekenstein bound the entropy
of the whole system  is $2\pi M_{MSH}R_{2}$. Regarding the Bekenstein bound
to reach maximum entropy, we have to consider the amount of $M_{MSH}$ by $R=R_{2}$. In doing so, by using the equations ($\ref{msh}$) and ($\ref{max}$) we have
\begin{equation}
M_{MSH}= \frac{H^{2}R^{3}}{2}+m_{0}=\frac{R}{2}.
\end{equation} 
Now we derive $m_{0}$ as follows
\begin{equation}\label{33}
2m_{0}=R^{2}(1-H^{2}R^{2}).
\end{equation}
Then, because of $m_{0}\geq 0$ we have $R\leqslant \frac{1}{H}$. 
Note that we obtained the equation (\ref{33}) in the sense that the local inhomogeneity,
namely $m_{0}$ part, is
a black hole.  If the  $m_{0}$ part is not a black hole, we still have
the 
same constraint $R\leqslant \frac{1}{H}$. Therefore, by imposing the
holographic principle we obtain a constraint over cosmological horizon by
Hubble radius.

We summarize our findings in this section in the following
remark. \\

\textbf{Remark 3:} \textit{When the local matter
system in the McVittie spacetime is a true black hole, the entropy-area
law  and holographic principle put two different constraints on the cosmological
horizon radius $R$. These constraints are the results of the fact that a black hole admits the saturated entropy. The first one states that there is a correspondence between the inner apparent horizon area and the cosmological horizon area which results in that the Hubble radius is discrete such that for any arbitrary value of black hole mass $m_{0}$, the value of $R$
is given up to an integer number. The latter represents that when a black
hole is placed in a cosmological background, the cosmological horizon radius
is constrained as $R\leq\frac{1}{H}$. The equality case is provided
only by disappearance of the black hole.}
\\
\section{Conclusion}

 Based on the generalized second law of thermodynamics, we have imposed a hypothesis of ``$D$-bound-Bekenstein bound identification'' for dilute
systems on the McVittie spacetime
surrounded by dark energy fields. We have found that \\  
\begin{itemize}
\item This identification is realized  for ``all times'' of cosmological constant domination $\omega=-1,$   ``early times'' of phantom field domination $\omega<-1$, and ``late times'' of quintessence field domination $\omega>-1$. \item This identification 
is
not realized for $\omega \neq -1$ in ``late times'' and ``early times'' for ``phantom''
and ``quintessence'' fields, respectively. 
\end{itemize}
Therefore, since the cosmological constant preserves
and respects this identification at ``all times'', we can single out the   cosmological
constant as the viable dark energy, consistent with this hypothesis, and rule out the   quintessence and phantom fields in the
study of McVittie spacetime. This result may have important impact on the
validity of the  studies on the McVittie spacetime in the
presence of ``phantom''
and ``quintessence'' fields.
  
 We have used discrete behaviour of entropy-area law  to derive constraint on the radius
 of cosmological horizon in McVittie solution
surrounded by cosmological fields.
These constraints reveal   a discrete feature  of the radius
 of cosmological horizon. The prominent point here is that all of these considerations are valid when the system can be  diluted. Regarding
the McVittie
solution surrounded by cosmological constant field, the system can be diluted, and the radius of
Hubble horizon can be treated in a discrete way according to the  discrete behaviour
of inner apparent horizon. This result is true for the McVittie solution with phantom field at early times and
quintessence field at  late times. At early times for quintessence field
and late
times for phantom field, the system cannot be diluted, and
the discrete behaviour of the inner apparent horizon does not manifest itself in the
discrete radius of Hubble horizon. 
\section{Acknowledgement}
We appreciate the anonymous referee
for  the constructive comments that significantly improved the quality of the paper. \textcolor[rgb]{1,0,0.501961}{
 }


\end{document}